\documentclass[aps,prl,showpacs,twocolumn,superscriptaddress]{revtex4} 
\usepackage{graphicx} 
\usepackage{amsmath} 
\usepackage{amssymb} 
\bibstyle{apsrev.bib} 
 
\newcommand{\beq}{\begin{equation}} 
\newcommand{\eeq}{\end{equation}}

\begin{document} 
\title{A Self-Organized-Criticality model consistent with statistical properties 
of edge turbulence in a fusion plasma} 
\date{\today} 
\author{Fabio Sattin} 
\affiliation{Consorzio RFX, ENEA-Euratom Association, 
Corso Stati Uniti 4, 35127 Padova, Italy} 
\email{fabio.sattin@igi.cnr.it}
\author{Marco Baiesi} 
\affiliation{Instituut voor Theoretische Fysica, K.U.Leuven, 
Celestijnenlaan 200D, 3001 Leuven, Belgium} 
\email{marco.baiesi@fys.kuleuven.be}
\begin{abstract} 
The statistical properties of the intermittent signal generated by a recent 
model for self-organized-criticality (SOC) are examined. A successful 
comparison is made with previously published results of the equivalent 
quantities measured in the electrostatic 
turbulence at the edge of a fusion plasma. This result re-establishes 
SOC as a potential paradigm for transport in magnetic fusion devices, 
overriding shortcomings pointed out in earlier works 
[E. Spada, {\it et al }, Phys. Rev. Lett. {\bf 86}, 3032 (2001); 
V. Antoni, {\it et al }, Phys. Rev. Lett. {\bf 87}, 045001 (2001)]. 
\end{abstract} 
\maketitle 
Self-organized criticality (SOC) is believed to account for the behaviour of 
several extended non-equilibrium systems exhibiting 
bursty activity with long-range correlations in space and time~\cite{0}. 
In the past years, SOC was candidated as 
a paradigm for the understanding of anomalous transport of energy and 
matter in magnetically confined fusion plasmas~\cite{1}. Within the huge 
complexity of transport in fusion devices, in fact, a number of 
features were identified, which could easily be cast into the framework 
of SOC systems: from the existence of critical average gradients and 
profile resilience, to the power-law power spectra of plasma 
parameters fluctuations (density, temperature, magnetic field, ...). 
Many properties of SOC numerical models are satisfactorily compared 
against experimental data \cite{2,3,6,7,8}: e.g.~transport barriers have been 
recently reproduced using SOC models~\cite{6}; non-locality, which is 
intrinsic in SOC, appears to be a possible ingredient for core 
transport~\cite{7}; numerical simulations of edge turbulence gave a 
phenomenology SOC-like~\cite{8}.

However, recently further analysis showed some 
irreducible discrepancies between numerical and experimental time series 
\cite{19,4,5}: these claims have by now confirmed in almost all devices. 
Essentially, numerical time series (normally coming from the 
Bak-Tang-Wiesenfeld (BTW) model~\cite{0}, 
the sandpile {\em \'a la} Kadanoff~\cite{9b}, or the Hwa-Kardar ``running'' 
sandpile~\cite{10}) were not found to display realistic temporal 
correlations. 
It was pointed out that experimental data display distribution 
of waiting times with power-law tails, and that intermittency 
appears as an ubiquitous property of plasma time series: a behaviour 
not registered in the analyzed SOC models. 
The first paper addressing the former point in the context of laboratory 
plasma was Ref.~\cite{4}, 
carrying on an analysis of the probability distribution function 
(PDF) of waiting times between bursts in density fluctuations 
measured on the RFX Reversed Field Pinch Experiment~\cite{11}. 
Experimentally, it was found a power-law curve with an 
exponent about $-2$, while the considered SOC 
models had exponential-like PDFs. 
Later, it has become clear that the statistics of waiting times 
is not truly a stringent test about the existence of SOC, because 
several SOC models have also non-exponential PDFs of waiting times 
(e.g., see~\cite{17,pbb} and references therein).

Intermittency in turbulence implies the lack of self-similarity 
between time scales~\cite{frisch}. 
It has been characterized in plasma experiments by looking 
at the PDF of signal differences (or with the most sophisticated tool of the 
continuous wavelet transform \cite{12}) at different time scales~\cite{5,13} . 
It turns out that the shape of the PDFs does not collapse to a single curve, 
irrespective of the time scale. It was found the presence of non-Gaussian PDFs 
in the RFX signal, approximately stretched exponentials, at the smaller $\tau$'s 
(higher frequency), gradually recovering the Gaussian shape as $\tau$ 
increased. Again, this is not the case in the analyzed SOC models.

SOC alone is a paradigm, which needs to be implemented into specific 
models in order to provide verifiable predictions. Part of the 
extracted predictions, hence, will be intrinsic to SOC, while others 
will depend on the model. 
Normally, as long as a specific model fails to account for some empirical 
evidence, one may devise a variant of it that is able to cure the specific 
shortcoming, but the improved accuracy of the final model is obtained 
at the expenses of some loss of generality. 
In our case, the simplicity 
of the SOC paradigm would be spoiled and obscured by model-specific 
features. A good option would be one where the added 
features reflect a property that the true physical system is anyway 
likely to possess.

Previously, turbulent features were already found in 
time series of ``waves'' in the BTW model~\cite{demenech}. 
The same features were later found in 
experimental time series of solar flares~\cite{bershadskii}. 
Waves are decompositions of avalaches, 
and by definition, within an avalanche, they start from the same site. 
Thus, a message from this fact is that spatial features of the drive are 
important, affecting the temporal correlations in the activity.

S\'anchez, Newman and Carreras advanced several suggestions, to make 
compatible the findings of Refs.~\cite{4,5} 
within a SOC-like scenario. The first 
proposal \cite{15} was that of a mixed avalanche-diffusive transport, 
where the self-similarity of timescales would be naturally broken by the 
second contribution. A later suggestion \cite{16} hinted to the possibility 
of inducing modifications to the waiting times distribution through 
correlations built into the drive, hence remaining within a pure SOC 
framework. However, the authors pointed out that introducing this 
sort of explicit correlations into a SOC model could be not 
trivial, since the generating mechanism is likely to be specific to 
any physical system, and therefore scarcely useful. We show that 
indeed it turns out to be a pessimistic view: a very simple and 
general correlation mechanism can still generate a very complex time 
series and output.

We are going to present in this work a SOC model 
that is able to reproduce the findings of papers~\cite{4,5}. This is 
obtained by postulating the existence of a correlation 
between successive locations of the fueling. This requisite appears 
rather plausible in a model that attempts to simulate the physics in a 
plasma device: particle and heat sources in these environments are 
more or less localized, in any case not uniformly randomly 
distributed.

\begin{figure}[!tb] 
\includegraphics[angle=0,width=8.0cm]{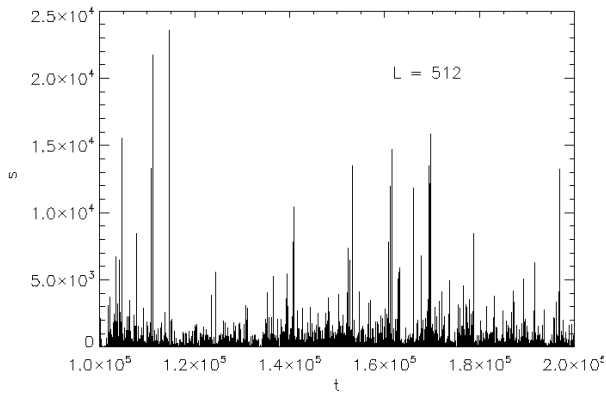} 
\caption{A sample of the time trace of the output of the model, 
showing its bursty behaviour. 
\label{fig:ts}} 
\end{figure} 

The SOC system is 
visualized as a standard 1-dim lattice automaton of length $L$, 
very similar to the sandpile studied by Kadanoff {\em et al}~\cite{9b}. 
The choice of the model and of the boundary conditions will allow 
a correspondence between its profile $h$ and ordinary profiles in 
laboratory plasmas. 
Each site $i$ $(1 \le i \le L)$ of the lattice holds $h_i$ ``grains''. 
Different choices 
for the boundary conditions can be taken: in \cite{17} periodic ones 
($h_{L+1} = h_1, h_0 = h_L$) were used. In this work, we will use one 
open boundary ($h_{L+1} = 0$), while the other one is closed 
($h_0 = h_1$). 
Stable pairs of adjacent sites $(j, \ell)$ 
fulfill the local stability condition if 
$|h_j-h_\ell|< H$, where the constant $H$ is a threshold. 
A local instability is resolved by a toppling, 
which consists in moving $\alpha$ 
grains from the upper to the lower site. 
This lower site in turn can become 
unstable, and chain reactions of transport are possible (avalanches). 
All instabilities in the systems are updated in parallel, and this procedure 
is iterated until the avalanche ends because all sites returned stable. 
The whole process takes place in one ``time step''. 
Instabilities arise because the system is driven out of equilibrium: 
a new grain is added at each time step at a position $i$: 
$h_i \to h_i + 1$. 
Correlations are introduced at the stage of choosing a new site $i'$ for 
deposition at time step $t$, provided that at the previous step ($t -1$) 
the grain had been deposited on site $i$. In earlier models~\cite{0,9b,10}, 
the choice is completely random and uncorrelated from the value 
$i$. Here, on the contrary, we make the opposite hypothesis of a strongly 
correlated diffusing dynamics: with equal probability, $i' = i + 1$ or 
$i' = i - 1$. Reflecting boundary conditions for the drive 
are used to avoid $i'$ to fall outside the lattice.

We do not expect qualitatively different results by varying slope-related 
quantities like $H$ and $\alpha$. 
Hence, we will keep them fixed from now on: $H = 4$, $\alpha=2$. 
We shall see, instead, that some results are not $L$-invariant. For reference, 
we will use $L = 512$. 
In what follows we will monitor for diagnostic purposes the total 
activity $s$, i.e.\ the total number of topplings within each avalanche. 
This activity is naturally identifiable with the quantity 
of matter or heat delivered to the device's walls, hence close to 
experimental investigation. 
All the simulations were carried out by starting with an empty system 
and letting it to load until a stationary critical phase is reached. 
Then, recording phases of several $ 10^6$ time steps were performed. 
One sample of the typical time series is given in Fig.~\ref{fig:ts}.

\begin{figure}[!tb] 
\includegraphics[angle=0,width=8.0cm]{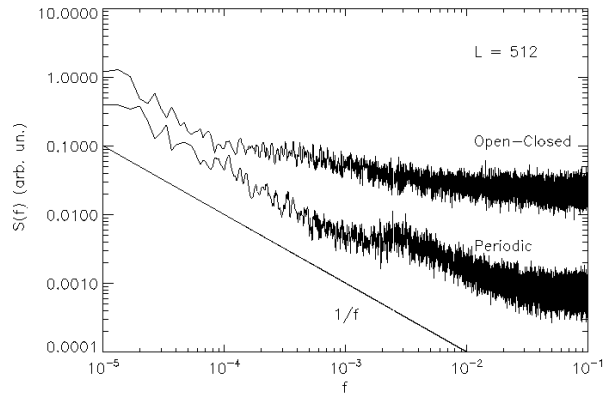} 
\caption{Power spectrum of the sinthetic signal for both periodic 
and open-closed boundary conditions. 
Overplotted for comparison is a $1/f$ curve. 
\label{fig:PowerSpectrum}} 
\end{figure} 
The power spectrum $S(f)$ is a statistical measure that can be easily 
computed from data sets, and gives fundamental information 
about time correlations built into the signal. 
Hence, an important issue is that about its scaling properties. The 
signature of SOC models is power-law spectrum. 
Most experiments (including also 
"numerical experiments") claim 
to find a power-law, or possibly more than one over different frequency ranges, 
with a variance of the exponent. Its absolute value, in several works, 
is higher than unity over most of the sampled frequency range, usually 
ranging between $2$ and $3$ \cite{20}. 
There are however some exceptions \cite{3,8,14,21} 
where a $1/f$ scaling appears to be 
recovered. It is possible that in the first set of papers the $1/f$ region 
exists too, but its width is too small to be detectable, and what was 
measured was instead just the high-frequency part of the spectrum (see, 
in this regard, Fig.~3 of Ref.~\cite{14}).

In Fig.~\ref{fig:PowerSpectrum} we show the power spectrum from the model, 
for two choices of boundary conditions (we study both because 
the choice of boundary conditions does affect the spectrum). 
In the periodic boundaries case, a $1/f$ slope 
is clearly discernible. In the other case, the trend still appears 
but lesser clear. 
This is quite remarkable a result: it is well known 
that the power spectrum of a random-walk-like signal is a $1/f^2$ curve. 
Hence, the random-walk dynamics of the driver couples to the rules 
for the stability of the sandpile to produce an highly nontrivial output.

\begin{figure}[!tb] 
\includegraphics[angle=0,width=8.0cm]{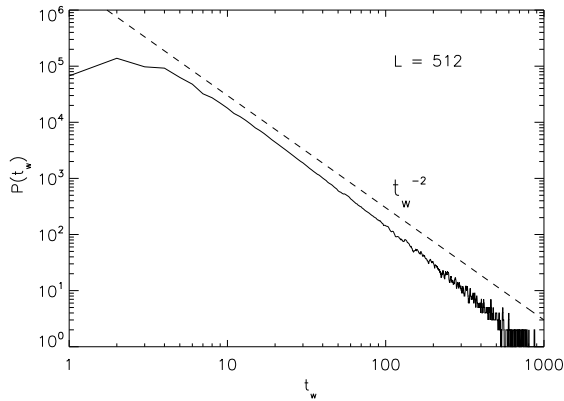} 
\caption{PDF of waiting times. Overplotted for comparison a $1/t_w^2$ 
curve. 
\label{fig:quietPDF}} 
\end{figure}

The first quantity of interest is the PDF of waiting times between 
bursts, $P(t_w)$, shown in Fig.~\ref{fig:quietPDF}. We 
define as bursts the points $s(t)$ for which 
$s(t) > 3 \times \langle s \rangle$, 
with $\langle s \rangle$ the average value over the sample. 
Most of the $P(t_w)$ curve is accurately fitted by 
a power-law with exponent $\approx -2$; interestingly, 
it is quite close to experimental values from RFX. 
The appearance of a non-Poissonian 
PDF is related to the existence of time correlations inherited 
by the driving~\cite{17}.

\begin{figure}[!tb] 
\includegraphics[angle=0,width=10cm]{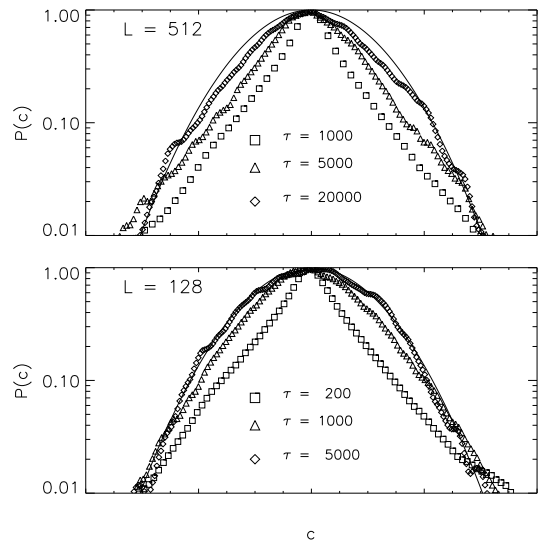} 
\caption{ 
Distributions of coefficient of wavelets transform at three time scales. 
The heights of the PDFs have been scaled to their maximum value, and 
the widths to the standard deviation. The solid curve is a Gaussian with 
the same amplitude and variance, plotted for reference. 
Upper plot, the lattice length is $L = 512$; lower plot, for comparison, $ L = 128$. 
\label{fig:diffPDF}} 
\end{figure}

So far, we have shown that the present model fulfills the questions 
raised in work \cite{4} as long as the scaling of waiting times is 
concerned. The issue of the departure from self-similarity is 
positively addressed in Fig.~\ref{fig:diffPDF}. There, we plot the PDF of the 
wavelet transform of the signal at three time scales. 
The PDF has fat tails at 
the smaller time scales, gradually approaching a Gaussian 
shape at the longest time scales. The explanation we give for the breaking of 
the self-similarity is this: the seeding has an effective "diffusivity" 
$D = $ (length step)${}^2$/(time step) = 1. After a time 
of order $ T = L^2/D = L^2$ the seeding has sampled the whole lattice, irrespective 
of the starting position. Thus, $T$ plays the role of a correlation time: 
the system keeps some memory of its past for as long as this time. 
Semi-quantitatively, some confirmations of this statement may be found: 
I) in figure~\ref{fig:diffPDF} we show that the convergence to Gaussian shape 
is faster with decreasing $L$. II) By sampling the system at a frequency 
$ f < 1/T$ one is actually sampling from independent realizations of the same system 
(Gibbsian ensemble). Hence, the scalar signal picked out 
must be equivalent to a random number from within a uniform finite distribution, 
whose power spectrum is white noise, $ f^0$. Indeed, the power spectrum in
fig.~\ref{fig:PowerSpectrum} begins to flatten towards an $f^0$ spectrum for 
$ f$ lesser than about $10^{-5}$, consistent with the $1/L^2 \approx 4\times 10^{-6}$ estimate.

Although successful when benchmarked against all the sought statistical tests, 
the model here proposed is still quite minimal, and add-ons may be envisaged 
that can enrich the built-in physics without appreciably spoiling 
its simplicity. An option we are pursuing is that of implementing into the model the 
running sandpile version of Hwa and Kardar~\cite{10}. Indeed, a time scale 
common to both the drive and the system dynamics appears a suitable feature for 
models of turbulent phenomena~\cite{pbb,1,6}.

Nowadays, the intermittent character of turbulence at the edge of 
fusion devices is attributed to the existence of long-lived coherent 
structures moving on top of a background plasma (dubbed blobs), which are 
responsible for most of the transport to the wall~\cite{coherent}. 
The precise mechanism of blob generation is self-regulating and 
involves the triggering of an interchange instability driven by 
pressure gradients with formation of a radially elongated structure, that ultimately 
is separated from the core plasma by the differential rotation and expelled 
towards the edge, with an accompanying relaxation of the mean profile \cite{8,blobs}. 
The affinities with SOC dynamics are apparent, and have been 
noticed by some authors (see, e.g.~\cite{gc,labombard}); however, to our knowledge,
the understanding of blobs' physics has not yet advanced enough to allow for
quantitative comparisons with SOC models.  

Undoubdtedly, the overall 
edge transport comes from an intricate pattern where the physics of 
creation, destruction, motion, and interaction of the coherent 
structures between them and with the background, needs to be carefully 
taken into account in order to provide a correct simulation of 
experimental evidence. It is therefore still an open question whether 
the SOC paradigm would be comprehensive enough to account for all this 
physics. The evidence presented in this paper suggests, however, 
that further analysis is necessary before discarding SOC as a 
potential transport paradigm. Indeed, in the past there has been 
sometimes a tendency to see a dichotomy between SOC and 
magnetohydrodynamics (MHD) turbulence that, perhaps, is there not: 
attempts of developing cellular automata consistent with MHD and 
Maxwell's equations are actually found in literature \cite{MHDandSOC}. 
\acknowledgements 
R. Cavazzana kindly provided the routine for wavelets analysis.
F.S.\ was supported by the European Communities under the contract of Association between Euratom/ENEA.

\end{document}